\documentclass[a4paper,12pt]{amsart}

\usepackage{amsmath}
\usepackage{amssymb}
\usepackage{amsfonts}
\usepackage{amsthm}
\usepackage[pagewise, mathlines]{lineno}

\newtheorem{theorem}{Theorem}

\newtheorem{proposition}{Proposition}

\DeclareMathOperator{\codim}{codim}

\newcommand{\entr}{s}
\newcommand{\vel}{u}
\newcommand{\temp}{\theta}
\newcommand{\dens}{\rho}
\newcommand{\visc}{\zeta}
\newcommand{\cond}{\kappa}
\newcommand{\press}{p}
\newcommand{\grav}{\mathrm{g}}
\newcommand{\PP}{\Phi}

\newcommand{\totalDiff}[1]{\frac{\mathrm{d}}{\mathrm{d}{#1}}}
\newcommand{\Jet}[1]{\mathbf{J}^{#1}}
\newcommand{\Eqtn}[1]{\mathcal{E}_{#1}}
\newcommand{\Clos}[1]{\widetilde{#1}}

\newcommand{\Point}{\mathrm{Point}}
\newcommand{\sym}{\mathrm{Sym}}
\newcommand{\smbl}{\mathrm{smbl}}
\newcommand{\s}{\xi}
\newcommand{\Lie}{\mathrm{Lie}}

\newcommand{\algg}{\mathfrak{g}}

\newcommand{\LieAlgebra}[1]{\mathfrak{#1}}
\newcommand{\Sdir}[1]{\partial_{#1}}

\title{Quotients of Navier--Stokes equation on space curves}

\author[A. Duyunova]{Anna Duyunova}
\address{V. A. Trapeznikov Institute of Control Sciences of RAS,
Profsoyuznaya street, 65, Moscow, Russia, 117997}
\email{anna.duyunova@yahoo.com}

\author[V. Lychagin]{Valentin Lychagin}
\address{V. A. Trapeznikov Institute of Control Sciences of RAS,
Profsoyuznaya street, 65, Moscow, Russia, 117997}
\email{valentin.lychagin@uit.no}

\author[S. Tychkov]{Sergey Tychkov}
\address{V. A. Trapeznikov Institute of Control Sciences of RAS,
Profsoyuznaya street, 65, Moscow, Russia, 117997}
\email{sergey.lab06@ya.ru}

\thanks{All three authors are partially supported by the RSF grant 21-71-20034}

\subjclass[2010]{35Q30}
\keywords{Navier--Stokes equation, quotient equation, exact solutions}

\begin{document}
\begin{abstract}
A Navier--Stokes system on a curve is discussed. The quotient equation for this system is found. The quotient is used to find some solutions of Navier--Stokes system. Using virial expansion of the Planck potential, we reduce the quotient equation to a series of systems of ordinary differential equations.
\end{abstract}
\maketitle
\section*{Introduction}
In the paper, a Navier--Stokes system on a specific class of curves is studied. Using its quotient equation for this system, we find some solutions of the Navier--Stokes system for different kinds of thermodynamic states.

The paper is organized as follows.

In Section \ref{sec:Quotients}, we briefly recollect the Rosenlicht and Lie--Tresse theorems, the notion of PDE quotients and relations between differential equations and their quotients.

In Section \ref{sec:NS}, we find a quotient equation for the Navier--Stokes system describing media flows on space curves in the field of constant gravity. Particularly, space curves with the $z$-component given by a linear function of curve length are considered. We use previous results \cite{Duyunova2020_2} regarding symmetry algebras and differential invariants, as well as their dependence on possible thermodynamic states and the form of a space curve. The field of differential invariants is generated by seven differential invariants and two invariant differentiations, and there are  five independent invariants of exact order $k\geq 2$. The latter leads us to the quotient PDE system of five equations of the first order. Also we study the problem of existence infinitesimal symmetries, or B\"{a}cklund-type symmetries, of the quotient. Noteworthy, the quotient system has two characteristic vector fields, though the original Navier--Stokes system has no characteristics.

In Section \ref{sec:sols}, we find solutions of the quotient using one of the characteristic vector fields. partially constant solution. In particular, solutions for the ideal and van der Waals gases are obtained. Also, we use virial expansion to rewrite the quotient as a series of ODE systems.

\section{Quotients of PDEs}\label{sec:Quotients}
\subsection{Algebraic structures in PDE geometry}
Let $\pi\colon E(\pi) \rightarrow M$ be a smooth bundle over a manifold $M$, $\dim \pi \geq 2$, and let $\pi_k \colon \Jet{k} \rightarrow M$, $k=0,1,\ldots,$ be the $k$-jet bundles of sections of the bundle $\pi$. The jet geometry is defined by the pseudogroup $\Point(\pi)$ of local diffeomorphisms of the manifold $\Jet{0}$. Bundles $\pi_{k,k-1}\colon\Jet{k}\rightarrow \Jet{k-1}$ are endowed with affine structures, which are invariant with respect to the Lie transformations, and prolongations of the pseudogroups  $\Point(\pi)$ are given by rational functions of $u_{\sigma}^i$ in the standard jet coordinates $\left( x, u_{\sigma}^i \right)$.

In other words, the fibers $\Jet{k,0}_{\theta}$ of the projections $\pi_{k,0} \colon \Jet{k} \rightarrow \Jet{0}$ at a point $\theta \in \Jet{0}$ are algebraic manifolds, and the stationary subgroup $\Point_{\theta}(\pi)\subset\Point(\pi)$ gives us birational isomorphisms of the manifold.

Differential equation $\Eqtn{k}\subset \Jet{k}$ is said to be {\it algebraic} if fibers $\Eqtn{k,\theta}$ of the projections $\pi_{k,0}\colon\Eqtn{k}\rightarrow\Jet{0}$ are algebraic manifolds.

We assume that differential equations here are formally integrable, and then all prolongations of an algebraic equation are algebraic too.

By a symmetry algebra of an algebraic differential equation, we mean the Lie algebra $\sym(\Eqtn{k})$ of point vector fields that acts transitively on $\Jet{0}$. Moreover, the stationary subalgebras $\sym_{\theta}(\Eqtn{k})$, where $\theta\in\Jet{0}$, or respectively $\theta\in\Jet{1}$, produce actions of algebraic Lie algebras on algebraic manifolds $\Eqtn{l,\theta}$, for all $l\geq k$.

\subsection{The Rosenlicht theorem}
Here we give the Rosenlicht theorem formulated for Lie algebras.

Let $B$ be an algebraic manifold, $\algg$ be a Lie subalgebra of the Lie algebra of the vector fields on $B$. By $\mathcal{F}(B)$ and $\mathcal{F}(B)^{\algg}$, we denote the field of rational functions on the manifold $B$ and the field of rational $\algg$-invariants on $B$ respectively.

Lie algebra $\algg$ is called {\it algebraic,} if there is an algebraic action of an algebraic group $G$ on $B$ such that $\algg$ coincides with the image of Lie algebra $\Lie(G)$ under this action. 
By an {\it algebraic closure} $\Clos{\algg}$ of a Lie algebra $\algg$, we mean the intersection of all algebraic Lie algebras that contain $\algg$.

We say that an orbit $Gb\subset B$ is {\it regular} (as well as a point $b$ itself) if there are $m=\codim Gb$ $G$-invariants $x_1,\ldots,x_m$ such that their differentials are linearly independent at the points of the orbit.

The Rosenlicht theorem states that the set of all regular point $B_0 \subset B$ is open and dense in $B$.

If the aforementioned invariants $x_1,\ldots,x_m$ are considered as local coordinates on the quotient $Q(B) = B_0/G$ (the set of all regular orbits at a point $Gb\in Q(B)$) then on the intersections of the coordinate charts the coordinates are connected by rational functions.

The field $\mathcal{F}(B)^{\algg}$ has transcendence degree equal to the codimension of regular $\Clos{\algg}$-orbits and it is also equal to the dimension of the quotient algebraic manifold $Q(B)$.

\subsection{Quotients of algebraic differential equations}
Let $\algg$ be an algebraic symmetry Lie algebra of an algebraic formally integrable differential equation $\Eqtn{k}$, and let $\Eqtn{l}$ be $(l-k)$-th prolongations of $\Eqtn{k}$. Then all equations $\Eqtn{l} \subset \Jet{l}$ are algebraic and we have the tower of algebraic bundles:
\[
\Eqtn{k} \longleftarrow \Eqtn{k+1} \longleftarrow \cdots
\longleftarrow \Eqtn{l} \longleftarrow \Eqtn{l+1} \longleftarrow \cdots.
\]

Let $\Eqtn{l}^0\subset\Eqtn{l}$ be the set of {\it strongly regular points} and $Q_l(\Eqtn{})$ be the set of all strongly regular $\algg$-orbits, where, by a strongly regular point (and orbit), we mean such points of $\Eqtn{l}$ that are regular with respect to $\algg$-action and their projections on $\Eqtn{l-1}$ are regular too.

Then $Q_l(\Eqtn{})$ are algebraic manifolds and the projections $\varkappa_l \colon \Eqtn{l}^0 \rightarrow Q_l(\Eqtn{})$ are rational maps such that the fields $\mathcal{F}\left( Q_l( \Eqtn{} ) \right)$ and $\mathcal{F}( \Eqtn{l}^0 )^{\algg}$, the field of rational functions on $\Eqtn{l}^0$ that are $\algg$-invariants \textit{(rational differential invariants),} coincide, i.~e., $\varkappa_l^{\ast}\left( \mathcal{F}\left( Q_l(\Eqtn{}) \right) \right) = \mathcal{F}( \Eqtn{l}^0)^{\algg}$.

The $\algg$-action preserves the Cartan distributions $C(\Eqtn{l})$ on the equations and, therefore, projections $\varkappa_l$ define distributions $C(Q_l)$ on the quotients $Q_l(\Eqtn{})$.


\begin{theorem}[The Lie--Tresse theorem]\cite{Kruglikov_2016}
Let $\Eqtn{k}\subset\Jet{k}$ be a formally integrable algebraic differential equation and let $\algg$ be an algebraic symmetry Lie algebra. Then there are rational differential $\algg$-invariants $a_1,\ldots,a_n,b^1,\ldots,b^N$ of order $\leq l$ such that the field of all rational differential $\algg$-invariants is generated by rational functions of these invariants and their Tresse derivatives $\frac{d^{\vert \alpha \vert } b^j}{da^{\alpha}}$.
\end{theorem}

Invariants $a_1,\ldots,a_n,b^1,\ldots,b^N$ are called Lie--Tresse coordinates.

In contrast to algebraic invariants, for which only the algebraic operations are possible, case of differential invariants provides us with an operation of taking Tresse derivatives that allows us to get actually new invariants.

Also, in case of differential invariants, syzygies provide us with new differential equations that are called {\it quotient equations.}

\subsection{Relations between differential equations and their quotients}
\begin{enumerate}
\item Let $u=f(x)$ be a solution of differential equation $\Eqtn{}$ and let $a_i(f)$, $b^j(f)$ be values of the invariants $a_i$, $b^j$ on the section $f$. Then, locally $b^j(f) = B^j\left( a(f) \right)$, and $b^j=B^j(a)$ is the solution of the quotient equation.
\item The correspondence between solutions is valid on the level of generalized solutions, in other words, on the level of integral manifolds of the Cartan distributions.
\item Suppose $b^j = B^j(a)$ is a solution of the quotient equation. Then, considering equations $b^j-B^j(a)=0$ as a differential constraint for the equation $\Eqtn{}$, we get a finite type equation $\Eqtn{}\cap\left\{ b^j-B^j(a) =0\right\}$ with a solution being a $\algg$-orbit of a solution of $\Eqtn{}$.
\item Symmetries of the quotient equation are B\"{a}cklund type transformations for the original equation $\Eqtn{}$.
\end{enumerate}

\section{Navier--Stokes equation on a curve}\label{sec:NS}
In this section we study the Navier--Stokes equation describing media flows on space curves in the field of constant gravity. We considered symmetry algebras and differential invariants, as well as their dependence on possible thermodynamic states and the form of a space curve, in~\cite{Duyunova2020_2}.

The Navier--Stokes system of PDEs describing such flows is the following \cite{Duyunova2020_2}
\begin{equation}\label{eq:NS1}
  \left\{
  \begin{aligned}
  &\dens(\vel_t  + \vel\vel_a)+ \press_a -\visc\vel_{aa}- \dens\grav h^{\prime}=0,\\
  &\dens_t + (\dens\vel)_a=0,\\
  &\dens\temp(\entr_t + \vel\entr_a) - \cond \temp_{aa}-\visc\vel_a^2=0,
  \end{aligned}
  \right.
\end{equation}
where $\vel$ are $\press$, $\dens$, $\entr$, $\temp$ are the flow velocity, pressure, density, specific entropy, temperature of the medium respectively, $\cond$ is the constant thermal conductivity, $\grav$ is the gravitational acceleration, $\visc$ is the medium viscosity and $h(a)$ is the $z$-component of a naturally-parametrised space curve.

Clearly, the system \eqref{eq:NS1} is incomplete, i.~e., it has five unknown functions and only three equations. The question of classification of possible thermodynamic relations is not considered here, since it was described in detail before \cite{Duyunova2020_1}. We assume these relation are given in terms of the Planck potential $\PP(\dens,\temp)$:
\[
\press(\dens,\temp)=-R\dens^2\temp\PP_{\dens},\quad
\entr(\dens,\temp)=R(\PP+\temp\PP_{\temp}),
\]
where $R$ is a specific gas constant. In particular, we consider the ideal gas law, i.~e.
\[
\PP=\frac{n}{2}\ln\temp-\ln\dens,
\]
where $n$ is the number of degrees of freedom of a medium particle.

As it was shown before \cite{Duyunova2020_2}, considering flows of a viscid medium on a space curve $M = \{x=f(a),\,  y=g(a), \, z=\lambda a\}$, in a field of constant gravitational field, we obtain the system
\begin{equation}\label{eq:NS2}
  \left\{
  \begin{aligned}
  &\dens(\vel_t  + \vel\vel_a)+ \press_a -\visc\vel_{aa}-\lambda\grav\dens=0,\\
  &\dens_t + (\dens\vel)_a=0,\\
  &\dens\temp(\entr_t + \vel\entr_a) - \cond \temp_{aa}-\visc\vel_a^2=0.
  \end{aligned}
  \right.
\end{equation}

To describe a symmetry Lie algebra of the system \eqref{eq:NS2} together with thermodynamic relations, we begin with consideration of a Lie algebra $\LieAlgebra{g}$ of point symmetries of the PDE system \eqref{eq:NS2} only.

Let $\vartheta \colon\LieAlgebra{g} \rightarrow \LieAlgebra{h}$ be the following Lie algebras homomorphism
\[
\vartheta\colon X\mapsto
X(\dens)\Sdir{\dens} + X(\entr)\Sdir{\entr} + X(\press)\Sdir{\press} + X(\temp)\Sdir{\temp},
\]
where $\LieAlgebra{h}$ is a Lie algebra generated by vector fields that act on the thermodynamic valuables $\press$, $\dens$, $\entr$ and $\temp$.

Recall \cite{Duyunova2020_1} that $\LieAlgebra{g}$ is generated by the vector fields 
\begin{align*} 
&X_1 = \Sdir{t},\quad X_2 = \Sdir{\press},\quad X_3 = \Sdir{\entr},\quad
X_4 = \Sdir{a},\quad X_5 = t\,\Sdir{a}+\Sdir{\vel},\\
&X_6 = t\,\Sdir{t}+2a\,\Sdir{a} + \vel\,\Sdir{\vel}-\press\,\Sdir{\press}-3\dens\,\Sdir{\dens}+2\temp\,\Sdir{\temp},\\
&X_7 = t\,\Sdir{t}+\left(\frac{\lambda\grav t^2}{2}+a\right)\Sdir{a}+\lambda \grav t\, \Sdir{u} - \press\,\Sdir{\press}-\dens\,\Sdir{\dens}.
\end{align*}

The pure thermodynamic part $\LieAlgebra{h_t}$ of the symmetry algebra is generated by the vector fields 
\begin{align*} 
Y_1 = \Sdir{\press}, \quad  Y_2 = \Sdir{\entr},\quad Y_3 = \press\,\Sdir{\press}+\dens\,\Sdir{\dens}, \quad Y_4 = \dens\,\Sdir{\dens}-\temp\,\Sdir{\temp}.
\end{align*} 

Thus, the Lie algebra of point symmetries of the Navier--Stokes system coincides with $\vartheta^{-1}(\LieAlgebra{h_t})$.

As we shown in \cite[Th.~2]{Duyunova2020_2}, if $h(a)=const$, or $h(a)=\lambda a$, the field of kinematic differential invariants are generated by the invariants:
\[
\dens,\quad
\temp,\quad
\vel_a,\quad
\dens_a,\quad
\temp_a,\quad
\vel_t+\vel\vel_a,\quad
\temp_t+\vel\temp_a
\]
and by the invariant differentiations
\[
\totalDiff{t}+\vel\totalDiff{a},\quad \totalDiff{a}. 
\]

\subsection{Quotient equation}
Choosing invariants $\dens$, $\temp$, $\vel_a$,
$\dens_a$, $\temp_a$, $\vel_t+\vel\vel_a$, $\temp_t+\vel\temp_a$ as Tresse coordinates $x$, $y$, $H_1$,  $H_2$, $H_3$, $H_4$, $H_5$ respectively, we get the quotient equation for \eqref{eq:NS2}
\begin{equation}\label{eq:QuotientNS2}
E_q=
\left\{
\begin{aligned}
& xy(xH_1 S_x - H_5 S_y) + \cond (H_2 H_{3x} + H_3 H_{3y}) + \visc H_1^2 =0,\\
& H_2 H_{1x} + \visc H_3 H_{1y} - H_2P_x - H_3P_y + x(\grav\lambda - H_4)=0,\\
& H_{2y}H_5 + x (H_{1y}H_3 - H_{2x} H_1) + H_2(xH_{1x} + 2H_1)=0,\\
& x H_1 H_{3x} - H_5 H_{3y} + H_2 H_{5x} + H_3 (H_{5y} - H_1)=0,\\
& x H_1 H_{1x} - H_5 H_{1y} + H_2 H_{4x} + H_3 H_{4y} - H_1^2=0,
\end{aligned}
\right.
\end{equation}
here the functions $P$ and $S$ are given by thermodynamic equations.

\subsection{Symmetries of the quotient}
In this section we study for which thermodynamic states, equivalently, functions $\PP$, the system \eqref{eq:QuotientNS2} has symmetries.
Direct computations show that this system has no symmetries if the function $\PP$ is arbitrary. Nevertheless, there are symmetries for some classes of $\PP$, two of which are presented below.
\begin{proposition}
If the Planck potential
\[
\PP(x,y)=C_1+C_2\ln y +\frac{1}{y}\left(C_3+\frac{C_4}{x}+C_5x^{\frac{\alpha_2}{\alpha_1}} \right),
\]
then the system \eqref{eq:QuotientNS2} admits a two-dimensional symmetry algebra
\[
\begin{split}
& \left\langle
\Sdir{y}, \alpha_1 x\Sdir{x} + \alpha_2 y \Sdir{y} + (\alpha_1+\alpha_2)H_1\Sdir{H_1}+
\left(2\alpha_1+\frac{\alpha_2}{2}\right)H_2\Sdir{H_2}+\right.\\
& \left.\left(\alpha_1+\frac{3}{2}\alpha_2\right)H_3\Sdir{H_3}+
\left(\alpha_1+\frac{3}{2}\alpha_2\right)(H_4-\grav\lambda)\Sdir{H_4}+
(\alpha_1+2\alpha_2)H_5\Sdir{H_5}\right\rangle,
\end{split}
\]
where $\alpha_1,\alpha_2, C_1,\ldots C_5$ are constants.
\end{proposition}

\begin{proposition}
If the Planck potential
\[
\PP(x,y)=f_1(y)+\left(C_1+\frac{C_2}{y}\right)\ln x + \frac{C_3}{xy},
\]
then the system \eqref{eq:QuotientNS2} admits a symmetry of the form
\[
x\Sdir{x}+H_1\Sdir{H_1}+2H_2\Sdir{H_2}+H_3\Sdir{H_3}+(H_4-\lambda\grav)\Sdir{H_4}+H_5\Sdir{H_5}
\]
where $C_1,C_2, C_3$ are constants, and $f_1$ is an arbitrary smooth function.
\end{proposition}

\section{Solutions of the quotient}\label{sec:sols}
In this section we find solutions of the quotient. Particularly, we look for solutions that are invariant with respect to the characteristics of the quotient.

To find the characteristics we calculate the symbol of \eqref{eq:QuotientNS2}, $\smbl(E_q)$, is
\[
\begin{pmatrix}
0 & 0 & \cond(H_2\s_1+H_3\s_2) & 0 & 0\\
\visc(H_2\s_1+H_3\s_2) & 0 & 0 & 0 & 0\\
x(H_2\s_1 + H_3\s_2) & -xH_1\s_1 + H_5\s_2 & 0 & 0 & 0\\
0 & 0 & xH_1\s_1 - H_5\s_2 & 0 & H_2\s_1 + H_3\s_2\\
xH_1\s_1 - H_5\s_2 & 0 & 0 & H_2\s_1 + H_3\s_2 & 0
\end{pmatrix},
\]
and its determinant
\[
\det \smbl(E_q) = \cond\visc (H_2\s_1 + H_3\s_2)^4(xH_1\s_1 - H_5 \s_2).
\]

The latter gives us two characteristic vector fields $Z_1=H_2\Sdir{x}+H_3\Sdir{y}$ and $Z_2=xH_1\Sdir{x} - H_5\Sdir{y}$.

It should be noted that, though the Navier--Stokes system \eqref{eq:NS2} does not have characteristics, its quotient \eqref{eq:QuotientNS2} has two: $Z_1$ and $Z_2$.

We find solutions of \eqref{eq:QuotientNS2} such that the functions $H_2$ and $H_3$ are first integrals of $Z_1$. Thus, solving the system $E_q \cup \{ Z_1(H_2)=Z_1(H_3)=0\}$, we obtain several classes of possible solutions of $E_q$.
\begin{enumerate}
\item For the case of ideal gas,
\begin{equation}\label{eq:qig}
\begin{aligned}
& H_1 = k_1 x,\quad H_2 = 0,\quad H_3 = k_2 x^\frac{n + 2}{n},\\
& H_4 = \grav\lambda- k_2 R x^\frac{n + 2}{n},\quad H_5 = 2 k_1x\frac{k_1\visc-Ry}{Rn},
\end{aligned}
\end{equation}
where $k_1$ and $k_2$ are constants.
\item For the van der Waals gas \cite{Lychagin2019}, i.~e.,
\[
\PP = \frac{n}{2}\ln y + \ln\left(\frac{3}{x} - 1\right) + \frac{9x}{8y},
\]
we have
\begin{align*}
& H_1 = c_2x,\quad H_2 = 0,\quad H_3=c_1x\left(\frac{x}{x-3}\right)^{\frac{2}{n}},\\
& H_4 = \grav\lambda+3c_1R\left(\frac{x}{x-3}\right)^{1+\frac{2}{n}},\quad
H_5 = 2 c_2 x\frac{ c_2\visc(x-3) + 3 R y}{nR(x-3)}.
\end{align*}
\end{enumerate}

Solutions for the case of ideal gas, we obtain by solving the system \eqref{eq:NS2} together with the relations \eqref{eq:qig}:
\begin{align*}
& \vel(t,a)=\frac{c_1 a}{c_1 t + c_2} + \frac{\grav\lambda t(c_1 t + 2 c_2) + c_3}{2(c_1 t + c_2)},\\
& \dens(t,a) = \frac{c_1 a}{c_1 t + c_2},\quad \temp(t,a)=\frac{c_4}{ (c_1t+c_2)^{\frac 2 n}} +
\frac {c_1\visc} R,
\end{align*}
where $c_1$, $c_2$, $c_3$ are constants.

Studying the case of van der Waals gas in the same manner, we get 
\begin{equation}\label{eq:vwsol}
\begin{aligned}
& \dens(t,a) = \frac{1}{c_2 t + c_3},\quad \vel(t,a) = \frac{c_2 a + c_4}{c_2 t + c_3} + f_1(t), \\
& \temp(t,a) = \frac{c_1 a}{c_2 t + c_3}(1 - 3 c_2 t - 3 c_3)^{-\frac 2 n} + f_2(t),
\end{aligned}
\end{equation}
where
\[
f_1(t)=\frac{R c_1 n (1 - 3 c_2 t - 3 c_3)^{-\frac{2}{n}} (n - 6 (c_2 t + c_3))
+ 3 \grav\lambda c_2 t (c_2 n t - 2  (c_2 t - c_3 (n-2)))}{6 c_2 (c_2 t + c_3) (n-2)},
\]
and $f_2$ satisfies the equation
\[
\begin{split}
& 9 (2-n) c_2 (c_2 t+c_3) ((1-3 c_2 t - 3 c_3) R n (c_2 t + c_3) f_2^{\prime} +\\
& 2 c_2 (3 R (c_2 t+c_3) f_2 + \visc c_2 (1- 3 c_2 t - 3 c_3))) +\\
& 3 (2-n) n c_1 c_2 R ((c_2 t + 2 c_3) \grav \lambda t + 2 c_4) (1 - 3c_2 t - 3 c_3)^{1-\frac 2 n} -\\
& R^2 n^2 c_1^2 (n - 6 (c_2 t + c_3)) (1-3 c_2 t-3 c_3)^{1-\frac 4 n} =0.
\end{split}
\]

\subsection{Virial expansion}
Let us make use of the fact that it is often convenient to consider the Planck potential $\PP$ in terms of so called{ \it virial expansion:}
\[
\PP(x,y)=\frac{n}{2}\ln y - \ln x - \sum_{i=1}^{\infty}\frac{x^i}{i}A_i(y).
\]
Thus, we can find solutions of the system \eqref{eq:QuotientNS2} in the form of power series of $x$:
\begin{align*}
&H_1(x,y)=x^{d_1}\sum_{k=0}H_{1,k}(y)x^k,\quad &H_2(x,y)=x^{d_2}\sum_{k=0}H_{2,k}(y)x^k,\\
&H_3(x,y)=x^{d_3}\sum_{k=0}H_{3,k}(y)x^k,\quad &H_4(x,y)=x^{d_4}\sum_{k=0}H_{4,k}(y)x^k,\\
&H_5(x,y)=x^{d_5}\sum_{k=0}H_{5,k}(y)x^k,
\end{align*}
where $d_1,\ldots,d_5$ are the integer constants that should be chosen such that \eqref{eq:QuotientNS2} can be expanded as power series of $x$. It can be shown that $d_1=2$, $d_2=1$, $d_3=0$, $d_4=4$, $d_5=2$. Hence, the zeroth order term of this expansion is a system of ordinary differential equations:
\begin{equation}\label{eq:expansion0}
\left\{
\begin{aligned}
& H_{3,0}^{\prime} = 0,\\
& RyH_{2,0} + RH_{3,0} - \grav\lambda = 0,\\
& H_{5,0} H_{2,0}^{\prime} + H_{3,0} H_{1,0}^{\prime} + 3H_{1,0}H_{2,0} = 0,\\
& H_{3,0} H_{5,0}^{\prime}-H_{5,0}H_{3,0}^{\prime} + 2H_{5,0}H_{2,0} - H_{3,0} H_{1,0}=0,\\
& H_{3,0} H_{4,0}^{\prime}-H_{5,0}H_{1,0}^{\prime} + H_{1,0}^2 + 4H_{2,0}H_{4,0}=0.
\end{aligned}
\right.
\end{equation}
Equations \eqref{eq:expansion0} can be solved straightforwardly:
\begin{align*}
& H_{1,0}=y^{2-\frac{2\lambda\grav}{c_1R}}\left(c_2 y^{-\frac{\lambda\grav}{c_1R}}+
c_3 \right), \quad
H_{2,0}=\frac{\grav\lambda-Rc_1}{Ry},\quad H_{3,0}=c_1,\\
& H_{4,0}=\frac{y^{4\frac{Rc_1-\grav\lambda}{Rc_1}}}{c_1}
\left(
\frac{c_2y^{\frac{Rc_1-\grav\lambda}{Rc_1}}}{Rc_1-\grav\lambda}
\left(
Rc_1c_2y^{\frac{-\grav\lambda}{Rc_1}}+c_3(2Rc_1-3\grav\lambda)
\right)+\right. \\
& \left. \frac{c_3(Rc_1c_3-2\grav\lambda)y+Rc_1^2c_4}{Rc_1}
\right),\\
& H_{5,0}=y^{3-\frac{2\grav\lambda}{Rc_1}} \left( c_3 +
\frac{Rc_1c_2}{Rc_1-\grav\lambda}y^{-\frac{\grav\lambda}{Rc_1}} \right),
\end{align*}
where $c_1,\ldots,c_4$ are arbitrary constants.

We can also give the system of equations for the first order term of asymptotic expansion
\begin{equation}\label{eq:expansion1}
\left\{
\begin{aligned}
& H_{3,0}H_{3,1}^{\prime} + (H_{2,0} + H_{3,0}^{\prime}) H_{3,1} = 0,\\
& R y (H_{3,0}A_1^{\prime} + 2 A_1 H_{2,0} + H_{2,1}) + R(H_{3,0}A_1 - H_{3,1}) - \visc(H_{3,0} H_{1,0}^{\prime} + 2 H_{1,0} H_{2,0}) = 0,\\
& H_{3,0} H_{1,1}^{\prime} + H_{5,0}H_{2,1}^{\prime} + 4 H_{2,0} H_{1,1} + 2 H_{1,0} H_{2,1} + H_{1,0}^{\prime} H_{3,1} +  H_{2,0}^{\prime} H_{5,1} = 0,\\
& H_{3,0} H_{5,1}^{\prime} - H_{3,0} H_{1,1} + 2 H_{5,0} H_{2,1} + (H_{5,0}^{\prime} - H_{5,0}) H_{3,1} + (3 H_{2,0} - H_{3,0}^{\prime}) H_{5,1} = 0,\\
& H_{5,0} H_{1,1}^{\prime} - H_{3,0} H_{4,1}^{\prime} - 3 H_{1,0} H_{1,1} - 4 H_{4,0} H_{2,1} - H_{4,0}^{\prime} H_{3,1} - 5 H_{2,0} H_{4,1} + H_{1,0}^{\prime} H_{5,1} = 0,
\end{aligned}
\right.
\end{equation}
solving which we obtain the first order terms of the expansion. Continuing along these lines, we obtain expansion terms $H_{i,j}$ up to the order we may need.


\begin{thebibliography}{9}
\bibitem{Duyunova2021} Duyunova, A., Lychagin, V., Tychkov, S., 2021. Quotients of Euler Equations on Space Curves. Symmetry 13, 186. https://doi.org/10.3390/sym13020186

\bibitem {Duyunova2020_2} Duyunova, A., Lychagin, V. and Tychkov, S., 2020. Symmetries and differential invariants for viscid flows on a curve. Global and Stochastic Analysis, 7(2), pp. 157-167.

\bibitem{Duyunova2020_1} Duyunova, A., Lychagin, V. and Tychkov, S., 2020. Symmetries and differential invariants for inviscid flows on a curve. Lobachevskii Journal of Mathematics, 41(12), pp.2435-2447.

\bibitem{Lychagin2019} Lychagin, V., Roop, M., 2019. Phase transitions in filtration of real gases,  arXiv preprint arXiv:1903.00276.

\bibitem{Lychagin2020} Lychagin, V., Roop, M., 2020. Critical Phenomena in Filtration Processes of Real Gases. Lobachevskii Journal of Mathematics, 41(3), pp.382–399.

\bibitem{krasil1996geometry} Krasilchchik, I., Vinogradov, A.M. and Lychagin, V.V., 1996. Geometry of jet spaces and nonlinear partial differential equations. Gordon and Breach.

\bibitem{Kruglikov_2016} Kruglikov, B. and Lychagin, V., 2016. Global Lie--Tresse theorem. Selecta Mathematica, 22(3), pp.1357-1411.

\bibitem{Rosenlicht_1956} Rosenlicht, M., 1956. Some basic theorems on algebraic groups. American Journal of Mathematics, 78(2), pp.401-443.

\bibitem{Tresse_1894} Tresse, A.R., 1894. Sur les invariants diff\'{e}rentiels des groupes continus de transformations. Acta mathematica, 18(1), p.1.
\end{thebibliography}
\end{document}